\documentclass[doublecol]{epl2} 

\usepackage[latin1]{inputenc}
\usepackage{graphicx,psfrag}
\usepackage{bm,longtable}

\usepackage{txfonts}

\newcommand{\E}{\mathrm{e}}

\newcommand{\average}[1]{\left\langle{#1}\right\rangle}

\newcommand{\p}[1]{\left({#1}\right)}
\newcommand{\pq}[1]{\left[{#1}\right]}

\title{Reconstructing the free energy landscape of a polyprotein by 
single-molecule experiments}
\author{A. Imparato\inst{1,2}\footnote{Present address:  ISI Foundation, Viale Settimio Severo 65, Villa Gualino, I-10133 Torino, Italy} \and F. Sbrana \inst{3} \and M. Vassalli \inst{3,4}}
\institute{
 \inst{1} Dipartimento di Fisica and CNISM, Politecnico di Torino, c. Duca degli Abruzzi 24, Torino, Italy\\
 \inst{2} INFN, Sezione di Torino, Torino, Italy\\
\inst{3} CSDC-Dipartimento di Fisica,Universit\`a di Firenze, via Sansone,1, Sesto Fiorentino, Italy\\
\inst{4} Istituto Sistemi Complessi, CNR, via Madonna del Piano 10, Sesto Fiorentino, Italy
}

\abstract{
The mechanical unfolding of an engineered protein composed of eight domains of Ig27 is investigated by using atomic force microscopy. Exploiting a fluctuation relation, the equilibrium free energy as a function of the molecule elongation is estimated from pulling experiments.
Such a free energy exhibits a regular shape that sets a typical unfolding length at zero force of the order of 20 nm. 
This length scale turns out to be much larger than the kinetic unfolding length
that is also estimated by analyzing the  typical rupture force of the molecule under dynamic loading.
}

\begin{document}
\maketitle
Force spectroscopy techniques have enormously increased our knowledge on the 
structure of biopolymers such as proteins and nucleic acids.
The possibility of controlling very precisely the force applied to the free ends of  such molecules allowed the experimental evaluation of the typical lengths and energies of the bonds stabilizing the molecular structures \cite{KSGB,rgo,cv1,DBBR,Ober1,Li_2000,exp_fc1,exp_fc2,NA_exp}.
The mechanical unfolding of a protein is typically an out--of--equilibrium process, where the unfolding 
occurs in time scales much shorter than the typical molecule relaxation time:
this prevents the possibility of performing the experiments in quasiequilibrium conditions and thus of obtaining direct  measurements of thermodynamic variables.  However, this problem can be overcome by using the remarkable equality derived by Jarzynski \cite{jarz},
which allows one to measure the free energy difference between the folded
and the unfolded state of a biomolecule \cite{jarzexp}. By exploiting
an extended version of the Jarzynski equality (JE) the free energy landscape of model biopolymers as a function of the molecular elongation has been evaluated
\cite{HumSza,seif, alb1, IPZ,IPZ1,ILT,LIT}.
Furthermore in a recent experimental work \cite{HSK}, the free energy landscape of a molecule similar to the one we use here has been reconstructed, 
in a range of the molecular elongation corresponding the unfolding of a single
domain (see below for a discussion of the protein structure).
Usually, the information concerning the landscape of a protein, obtained by single--molecule experiments, has
been limited to the position and to the height of the energy barrier along the reaction coordinate, which is usually the molecule elongation.
The actual shape of the landscape can be only guessed by adapting it to the information
about the kinetics gathered during the unzipping experiments, see e.g., \cite{SR}.
However, in a recent paper \cite{IPZ1} it has been argued that the unfolding length and the unfolding rate that are obtained by investigating the kinetics of  protein unfolding, are not simply related
to the position and height of free energy barriers as given by the extended JE.
In other words, the kinetic parameters governing the protein unfolding may not correspond to the {\it 
equilibrium} thermodynamical properties of the molecules.

The aim of this paper is thus to show that it is possible to reconstruct the free energy landscape (FEL)  of a  protein for any value of its elongation.
Furthermore, we aim to compare the typical length scale of this landscape with 
the unfolding length measured in kinetic experiments, and to discuss
the meaning of this parameter in the context of the equilibrium properties of proteins.

\begin{figure}[h]
\center
\psfrag{v}[ct][ct][1.5]{$v$}
\psfrag{z}[bb][bb][1.5]{$z$}
\psfrag{L}[ct][ct][1.5]{$\ell$}
\includegraphics[width=8cm,height=5cm]{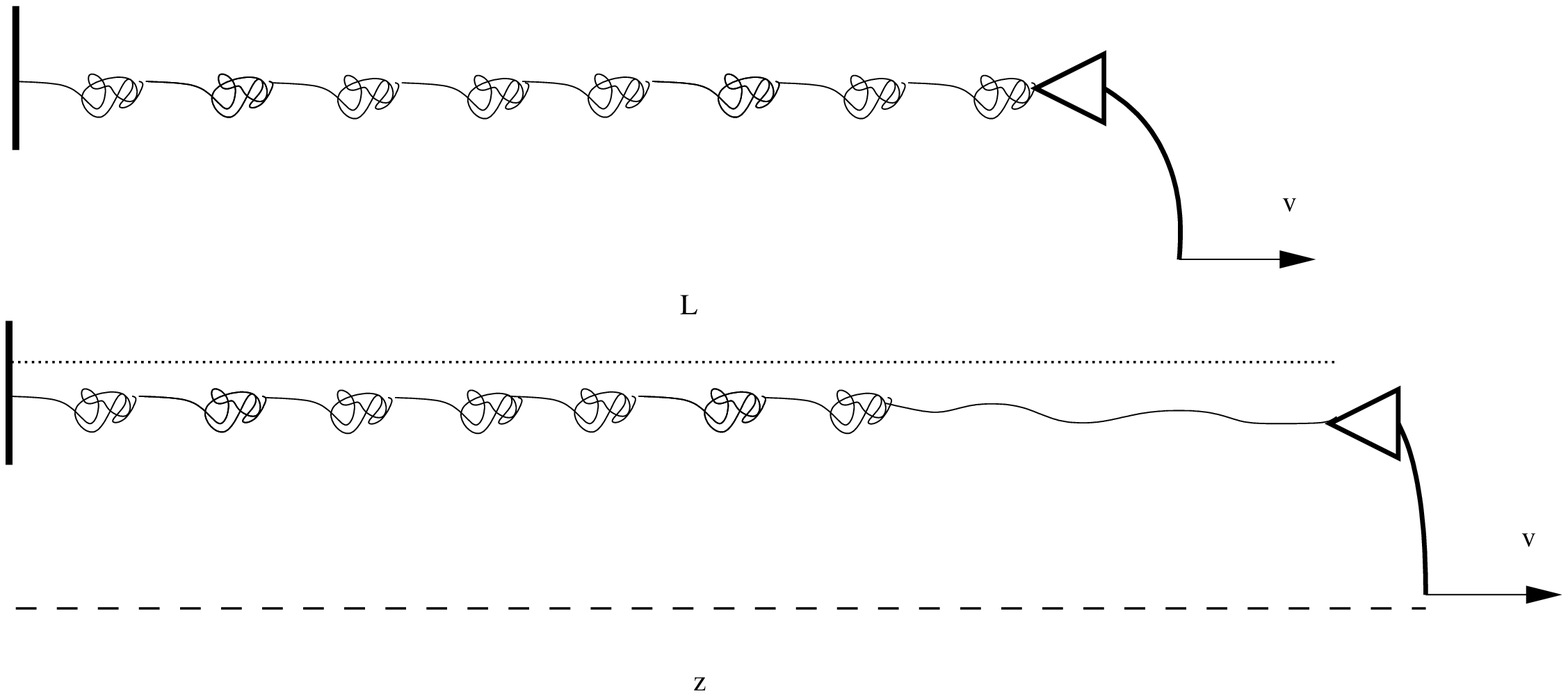}
\psfrag{z}[ct][ct][1.]{$z$ (nm)}
\psfrag{f}[ct][ct][1.]{$f$ (pN)}
\includegraphics[width=8cm]{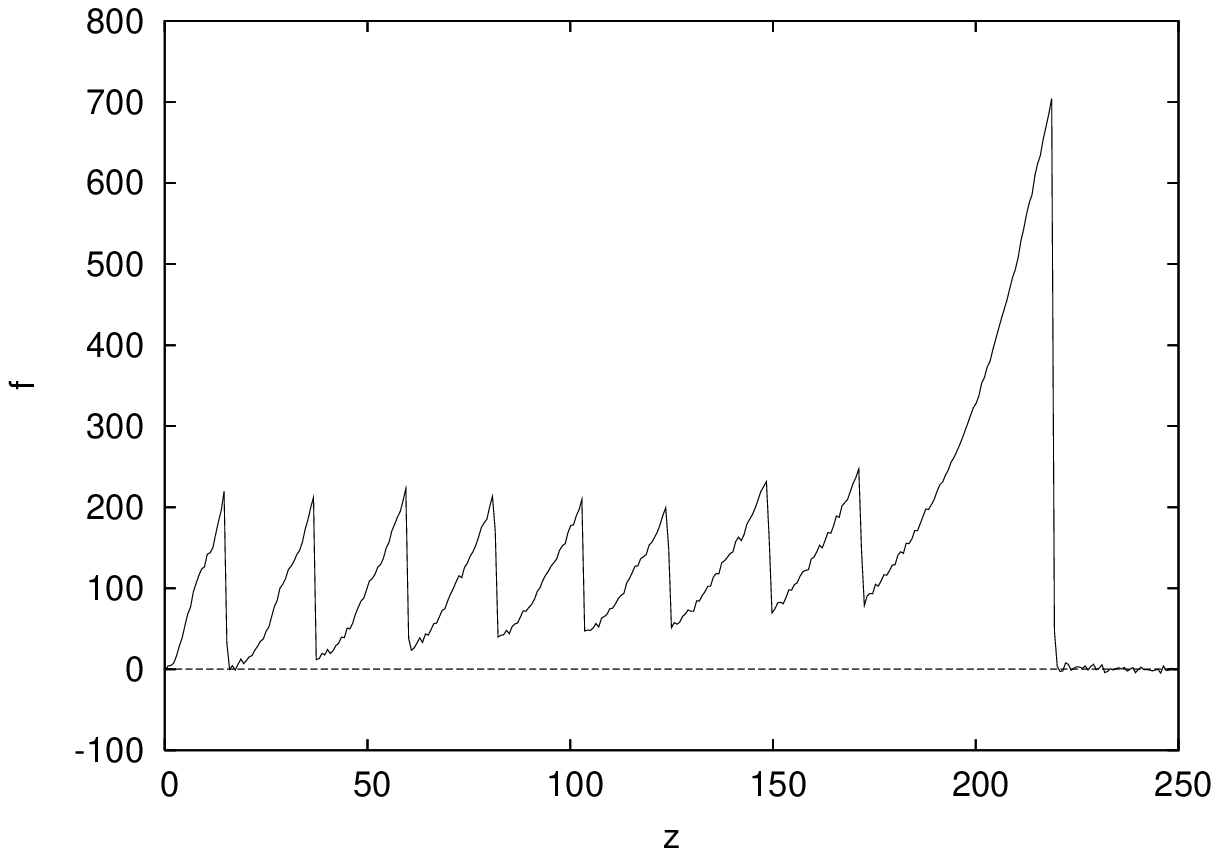}
\caption{Top panel: Cartoon of the protein attached to the tip on an AFM. Bottom panel: Force extension curve  of the polyprotein composed of
 eight repeats of the Ig27 domain.  The typical distance between two consecutive peaks ranges between 22 and 26 nm.} 
\label{saw}
\end{figure}

Unfolding of large proteins is a very rare and slow event, but by using single--molecule techniques it is possible to drive mechanical unfolding and make it feasible to study the unfolding kinetics of these molecules.
The molecule investigated in this work is a recombinant polyprotein composed of
 eight repeats of the Ig27 domain of human titin in PBS buffer (I270TM, Athena Enzyme Systems (TM)  Baltimore, MD, Cat. No.0304).
The proteins were allowed to absorb onto a gold substrate, and any free protein was removed by washing the sample off. In single--molecule force spectroscopy experiments exploiting atomic force microscope  (AFM), the molecule  deposited onto substrate is grabbed by the AFM probe. Mechanical unfolding of the protein is thus induced by moving the probe away from the substrate with a constant velocity $v$ (linear protocol). As the probe is retracted, the force exerted on the molecule increases until the molecule
 suddenly unfolds. If the molecule is composed of multiple domains, as in the present paper, further retraction causes the extension of successive folded domains, until all the domains 
are unfolded and the protein--tip interaction is broken. The corresponding 
force-distance curve has a typical sawtooth structure, as shown in fig.~\ref{saw}.
The probe of an AFM behaves as an harmonic spring within a wide range of values of the deformation with respect to the equilibrium position, and thus the force exerted on the molecule is $f=k (z(t)-\ell)$  where $k$ is the spring constant, $\ell$ is the elongation of the molecule and $z(t)=v t$ is the actual position of the probe with respect to the substrate (see fig.~\ref{saw}).
In the present work, mechanical unfolding
 experiments are performed with velocities $v$ ranging from 
200 to 2000 nm /s at room temperature.

\begin{figure}[h]
\center
\psfrag{f}[ct][ct][1.]{$f^*$ (pN)}
\psfrag{v}[ct][ct][1.]{$v$ (nm/s)}
\includegraphics[width=8cm]{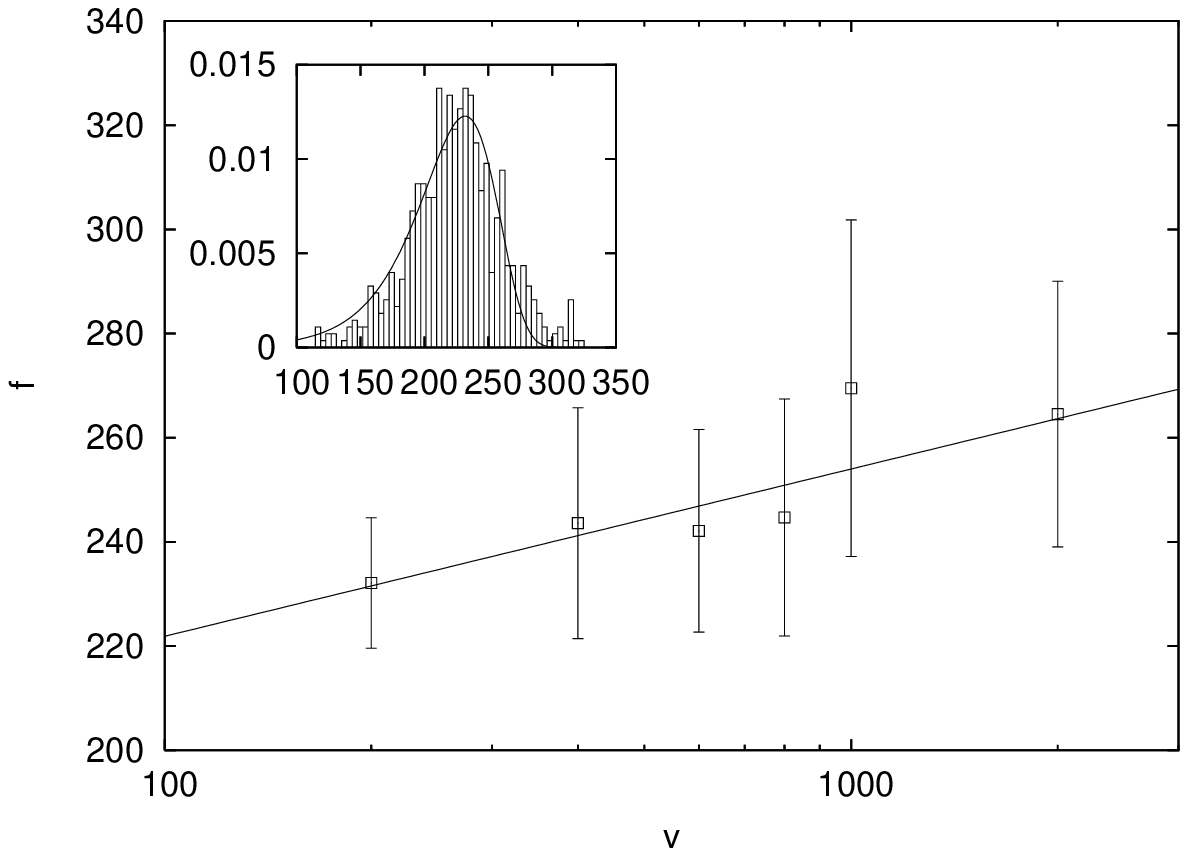}
\caption{Rupture force $f^*$ of a single Ig27 domain as a function of the AFM probe velocity $v$, the line is a fit to eq.~(\ref{fstar}), the number of single domain rupture ranges between 104 and 751 ($v=200 \mathrm{nm/s} \rightarrow751$ single rupture events, $v=400 \mathrm{nm/s} \rightarrow367$, $v=600 \mathrm{nm/s}\rightarrow209$, $v=800 \mathrm{nm/s}\rightarrow286$, $v=1000 \mathrm{nm/s} \rightarrow104$,  $v=2000  \mathrm{nm/s} \rightarrow351$). Inset:
Histogram of the rupture force for $v=200$ nm/s, the line is a fit to eq.~(\ref{p_f}).}
\label{v_f}
\end{figure}

Let us now consider the case where a molecular bond, or a group of molecular bonds are subject to 
a force increasing linearly with the time $f= r t$, where $r$ is the force increase rate. It can be shown that 
the rupture force is distributed according to the probability distribution function  \cite{ev2}
\begin{equation}
P(f)=\frac{1}{\tau_0 r} \E^{\beta f x_u} \exp\pq{-\frac{k_B T}{r x_u\tau_0} \p{\E^{\beta f x_u}-1}};
\label{p_f}
\end{equation} 
and the typical rupture force, defined as the maximum of $P(f)$,  is given by 
\begin{equation}
f^*=\frac {k_B T}{x_u}\ln\pq{\beta r  x_u \tau_0},
\label{fstar}
\end{equation} 
where $x_u$ is the unfolding length, usually interpreted as the deformation after which the molecule unfolds, 
$\tau_0$ is the typical unfolding time at zero force, and $\beta=1/(k_B T)$.
Since we use a rather soft spring ($k=0.04$ N/m), we take the rate of increase of the force in eqs.~(\ref{p_f})-(\ref{fstar}) to be $r=k v$. By plotting the force as a function of the time, 
we verified that this is a good approximation for the force rate (data not shown).
Because the AFM tip can bind any module of the protein
 at random, one is not able to
 control the site from which the protein is 
picked up \cite{Li_2000,exp_fc1}. Therefore, we obtained
a random sample of single molecule unfolding trajectories
containing from one to eight unfolding events.
In figure \ref{v_f} we plot the typical rupture force $f^*$  of a single Ig27 module as a function of the AFM probe velocity $v$, where trajectories with at least three repeats (i.e. rupture events) are considered.
By fitting the data to eq.~(\ref{fstar}) we obtain an estimate of the unfolding length $x_u=0.30\pm 0.07$ nm, which is in good agreement with the value $x_u=0.25$ nm found in \cite{rgo}.

\begin{figure}[h]
\center
\psfrag{F}[ct][ct][1.]{$F_0\, (k_B T)$}
\psfrag{L}[ct][ct][1.]{$\ell$ (nm)}
\includegraphics[width=8cm]{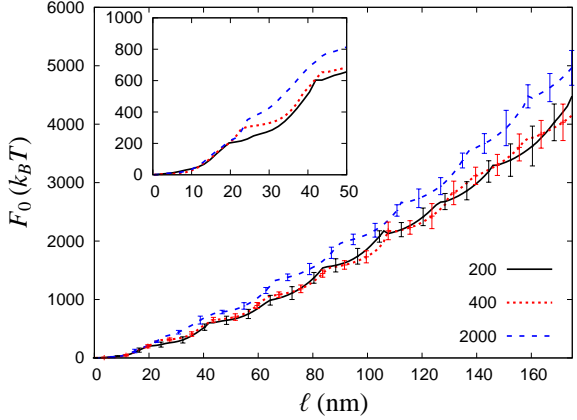}
\caption{Free energy landscape $F_0$ as a function of the molecular elongation $\ell$, as given by eq.~(\ref{je}) for different values probe velocity, $v=200,\, 400,\, 2000$ nm/s: the number of unfolding trajectories considered are 66, 35, 29, respectively. The error bars are obtained by using the jackknife approach \cite{Ef} for data resampling: for each value of $v$ subsamples of the total number of trajectories are considered, and for each subsample the free energy landscape is evaluated, as discussed in the text. One obtains thus a set of curves $F_0(\ell)$, and  the error bars are obtained, for each value of $\ell$, as the standard deviation of the mean with respect to this set.  The bars are mutually shifted for clarity's sake. Inset: Plot of $F$ in the small $\ell$ range. }
\label{coll_fig}
\end{figure}

We now focus on the evaluation of the free energy landscape $F_0(\ell)$, which is a function of the molecular elongation $\ell$.
As discussed in ref.~\cite{HumSza}, such an energy landscape can be obtained by exploiting an extend version of the Jarzynski equality
\begin{equation}
  \average{\delta(\ell-\ell_t) \E^{-\beta W_t}}_t \E^{\beta U(\ell,z(t))}
  =\exp\pq{-\beta F_0(\ell)}/Z_0,
\label{je}
 \end{equation}
where $W$ is the work done on the molecule during the unfolding process,  
$Z_0$ is the partition function of the unperturbed system (folded molecule with $f=0$), and $U(\ell,z(t))=k/2 (z(t)-\ell)^2$ is the external potential associated to the tip of the AFM.
Note that, when one takes the logarithm of the rhs of eq.~(\ref{je}), the partition function appears in the additive constant $-\log Z_0$, and thus it plays no role in the setting of $F_0(\ell)$.
In the following, we do not consider all the unfolding trajectories that we used to obtain the data in fig.~(\ref{v_f}), but we take only those trajectories with at least six rupture events. For each of these trajectories, we compute the work $W$ done on the molecule by the external force $f(t)=k (z(t)-\ell(t))$.
Practical procedures for obtaining the optimal estimate of $F_0(\ell)$ from eq.~(\ref{je}) are discussed in refs.~\cite{HumSza,seif,alb1}.
However, as the value of the work $W$ exceeds few hundreds of $k_B T$, eq.~(\ref{je}) gives unreliable results, since in that case the average is estimated by 
considering very small numbers ($\exp(-\beta W)\ll1$).
In order to avoid this numerical problem,  we calculate here the  average of the quantity $\exp(-\beta W+\Delta)$, where $\Delta$ is a fixed quantity, so as eq.~(\ref{je}) provides an estimate of
$\exp\pq{-\beta \left(F_0(L)+\Delta\right)}$.
By taking different values for $\Delta$ one selects ranges of values  of the work $W$ such that $W+\Delta$ is not larger than a few hundreds of  $k_B T$, and thus one can reconstruct the free energy landscape $F_0(\ell)$ piecewise.
In particular we take $\Delta=0,\, 1000,\, 2000,\, 3000,\, 4000\, k_B T$.

In refs.~\cite{alb1,IPZ} it was argued that eq.~(\ref{je}) supplies a reliable estimate
of the molecule free energy, provided that, as  the pulling rate is decreased,
the estimated curves $F_0(\ell)$ collapse onto a single curve.
In figure \ref{coll_fig}, the free energy landscape $F_0(\ell)$ as reconstructed from eq.~(\ref{je}) is plotted for different values of the AFM tip velocity $v$.
We actually observe a collapse of the curves for the two smallest velocities, while the agreement with the curve obtained with the largest value of $v=2000$ nm/s worsens as the coordinate $\ell$ increases.
\begin{figure}[h]
\center
\psfrag{F}[ct][ct][1.]{$F\, (k_B T)$}
\psfrag{L}[ct][ct][1.]{$\ell$ (nm)}
\psfrag{f=0}[lt][lt][.7]{$f=0$ pN}
\psfrag{f=50}[lt][lt][.7]{$f=50$ pN}
\psfrag{f=100}[lt][lt][.7]{$f=100$ pN}
\includegraphics[width=8cm]{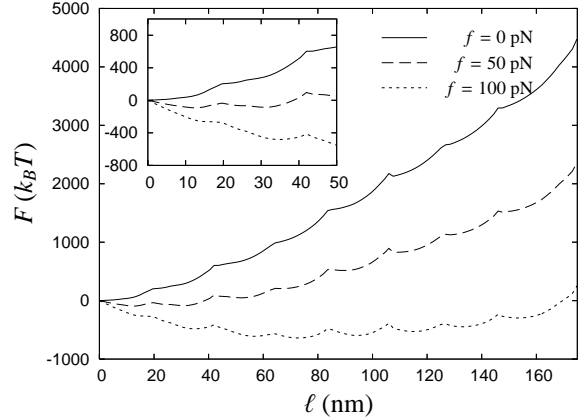}
\caption{Free energy landscape $F(\ell,f)=F_0(\ell)-f \ell$ for three values of the force, $f=0,50, 100$ pN. The  function  $F_0(\ell)$ corresponds to the most reliable estimate of $F_0(\ell)$ we obtained, i.e. the one with $v=200$ nm/s.}
\label{land}
\end{figure}

The energy landscape $F_0(\ell)$ exhibits a single minimum at $\ell=0$, corresponding to the molecule in the folded native state.
However the curve $F_0(\ell)$ exhibits cusps which are equispaced at a distance of $\Delta \ell\simeq20$ nm.
As a constant force $f$ is applied, the free energy landscape is tilted according to
$F(\ell,f)=F_0(\ell)-f \ell$. In figure \ref{land} we plot the FEL, as obtained with the smallest velocity here considered,  for different values of the force $f$.
As the external force $f$ is applied, the landscape $F(\ell,f)$ becomes more intricate: the cusps become local
maxima, delimiting equispaced minima. 
The overcoming of each such cusp corresponds thus to the breaking of a single Ig27 domain.
For each value of the force, a well defined global minimum appears in the function $F(\ell,f)$, 
therefore  $F(\ell,f)$ predicts how many domains out of the eight will be unfolded at equilibrium for that value of the force.
As an example, we find that the second minimum becomes deeper that the first one as $f\ge 50$ pN, see inset of fig.~\ref{land}. Thus, we expect that in a force clamp experiment, where a feedback system allows to 
apply a constant force to the molecule free ends \cite{exp_fc1,exp_fc2}, this would be the constant force required to unfold a single Ig27 domain. In a different experimental set-up, where force-ramp pulling was used, the first Ig27 domain was found to unfold at a typical force 
$f\simeq 75$ pN (see fig. 5b of ref.~\cite{Ober1}): such a value is consistent with $f\simeq 50$ pN predicted by our results.

We want to stress that at zero force the distance between the global minimum $\ell=0$ and the first cusp is  $\Delta \ell(f=0)\simeq20$ nm, while at $f=50$ pN the distance between the first minimum and the following maximum is $\Delta \ell(f=50) \simeq 6$ nm. These lengths are one order of magnitude larger than  the kinetic unfolding length $x_u=0.25$ nm found in \cite{rgo} and in the present work. 
Thus as discussed in ref~\cite{IPZ1}, the quantity $x_u$ appearing in eq.~(\ref{fstar}) represents a kinetic parameter, and does not correspond to  the typical length  of the free energy landscape: at zero force the thermal fluctuations  have to induce a deformation $\Delta \ell=20 \mathrm{nm}\gg x_u$ before the molecule unfolds, while at constant force $f=50$ pN this deformation has to be  $\Delta \ell=6$ nm.

Let us now discuss the reliability of the reconstruction of $F_0(\ell)$ over the whole range of values of $\ell$.
As the AFM probe is retracted and the molecule progressively unfolds, the 
system finds itself farther and farther from equilibrium. Since the Jarzynski equality requires that all the trajectories start from the equilibrium condition \cite{jarz}, we are more confident on the reliability of the reconstruction of $F_0(\ell)$ in the small-to-moderate $\ell$ range, than in the large $\ell$ range.
However, inspection of  fig.~\ref{land} indicates that each of the eight energy wells exhibits an almost  identical shape. In fact the distance between two successive cusps is constant, and at $f>0$ the
height of each relative maximum with respect to the respective energy minimum
is also constant, within a few $k_BT$-s.

We now compare our results with those of ref.~\cite{HSK}, where the FEL of a similar molecule was investigated.
In that paper the reconstruction of the FEL was limited to $\ell<30$ nm, i.e. to the rupture of a single Ig27 domain, while here we manage to reconstruct the FEL for the whole range of the molecular extension, as discussed above.
In ref.~\cite{HSK} the minimal velocity used is $v=50$ nm/s, which is one fourth of the smallest velocity used here $v=200$ nm/s.
However for this velocity, we find that at the first cusp ($\ell^* \simeq20$ nm) the free energy takes the value $F_0(\ell^*)\simeq200\, k_B T$ (see inset of fig.~\ref{coll_fig}) which is slightly larger than the value found in that paper $F_0(\ell \simeq20)\simeq 167\, k_B T$.
This result indicates that although we use a larger velocity compared to ref.~\cite{HSK}, we obtain a reliable estimate of the FEL for the polyprotein considered, and using a smaller velocity would only give negligible corrections to our estimate.

We find that, 
at $v=200$ nm/s, the average value of the work performed during the unfolding of a single Ig27  domain is $\average{W^*}\simeq 405\,  k_B T$, where $W^*$ corresponds to  the value of the work, when the position of the probe is $z^*=\ell^*\simeq20$ nm. The average work to unfold the successive  Ig27  domains takes the same value within statistical errors.
This value has to be compared with the value of the energy landscape, 
$F_0(\ell^*)\simeq 200\, k_B T$,  i.e. for the value of the coordinate $\ell$ where a single domain is unfolded, see figure \ref{coll_fig}. 
Thus, we have $\average{W^*} > F_0(\ell^*)$, and can conclude that the Jarzynski equality gives a better estimate of $F_0(\ell)$ than the classical measurement of $\average{W}$.

In conclusion we have evaluated the free energy landscape of a polyprotein
by combining atomic force microscopy and the JE. 
The landscape exhibits equispaced wells, each corresponding to a single domain of the protein.
The typical length scale of this landscape is 20 nm, which is larger than the 
typical unfolding length $x_u$ as obtained by analyzing the typical rupture force. We want to remark that in ref. \cite{rgo} the length $x_u=0.25$ nm was identified as the distance of the free energy barrier from the folded state.  
Our results suggest, on the contrary, that, at $f=0$, the distance between the folded state  and the first energy barrier is of the order of 20 nm. We attribute this difference to the fact that the quantity $x_u$ as obtained by eq.~(\ref{fstar}) is a kinetic parameter describing the unfolding kinetics of the Ig27 domain, while 
the extended JE (\ref{je}) provides an equilibrium quantity, namely the equilibrium FEL.

It is worth to note that as our approach is successful to reconstruct 
the FEL of a protein made up of identical domains in the whole range of the molecule extension, we believe that the same approach can be successfully used to 
probe the FEL of heterogeneous proteins made up of different domains.

\acknowledgments
FS and MV acknowledge the support of the  Strep European Project n.~12835 EMBIO
 (EMergent organization in complex BIOmolecular system).

\end{document}